\documentclass[12pt]{article}
\newcommand{\be}{\begin{equation}}
\newcommand{\ee}{\end{equation}}
\newcommand{\bear}{\begin{eqnarray}}
\newcommand{\eear}{\end{eqnarray}}
\newcommand{\ba}{\begin{array}}
\newcommand{\ea}{\end{array}}

\newskip\humongous \humongous=0pt plus 1000pt minus 1000pt

\newif\ifdtup

\def\oldreffmt#1{\rlap{[#1]} \hbox to 2\parindent{}}

\def\figfmt#1{\rlap{Figure {#1}} \hbox to 1in{}}

\def\beq{\begin{equation}}
\def\eeq{\end{equation}}
\def\bea{\begin{eqnarray}}
\def\eea{\end{eqnarray}}
\def\half{\frac{1}{2}}

\def\bq{\begin{quote}}
\def\eq{\end{quote}}

\relax

\newdimen\tdim
\tdim=\unitlength
\def\bar{\overline}

\begin{document}

\pagestyle{empty}
\begin{titlepage}
\def\thepage {}    

\title{Physics of Extra Dimensions as
IR Limit of Four-Dimensional Theories\footnote{\bf Talk given at the IPPP
Workshop on the Phenomenology of Beyond the Standard Model, Durham,
Great Britain, 7-10 May  2001, and at Planck2001 ``From the Planck Scale
to the Electroweak Scale'', La Londe-les-Maures, France, 11-16 May 2001}   
 \\[2mm] } 
\author{\bf Stefan Pokorski$^{1}$ \\[2mm]
{\small {\it $^1$ Institute for Theoretical Physics}}\\
{\small{\it Hoza 69, 00-681, Warsaw, Poland}}}


\maketitle

\vspace*{-10.0cm}
\noindent
\begin{flushright}
IFT-01/22 \\ [1mm]
July, 2001
\end{flushright}

\vspace*{10.1cm}
\baselineskip=18pt

\begin{abstract}
I discuss the recent work done in collaboration with Chris Hill, Jing Wang
and Hsin-Chia Cheng. We construct four-dimensional renormalizable gauge 
theories which, in their infrared limit, generate  the dynamics of gauge
interactions  in flat extra dimensions.  In this construction,
the Kaluza-Klein states of extra dimensions arise from having many gauge
symmetries in four dimensions.  The IR limit is determined by the dynamics
of spontaneous breakdown of those symmetries.
\end{abstract}

\vfill
\end{titlepage}

\baselineskip=18pt
\renewcommand{\arraystretch}{1.5}
\pagestyle{plain}
\setcounter{page}{1}


%





Non-abelian gauge theories in 4+n dimensions have dimensionful coupling 
constant $g_0=[M]^{-n/2}$ and are non-renormalizable. They require a cut-off
at some scale $M_s$ .  Such theories should be viewed as the
effective low energy limit
of some more fundamental theories, with better UV behaviour.

It  is an interesting question how uniquely the physics of a gauge theory in
4+n dimensions defines its underlying UV ``completion''. One may expect
that there exist (universality) classes of theories that have identical
behaviour in the infrared but are different above the scale $M_s$.
For instance, it is possible that a given non-abelian gauge theory in
4+n dimensions has some stringy UV completion.  In ref.~\cite{hill} we 
demonstrate a more surprising fact: there exist renormalizable
four-dimensional gauge theories that in the infrared regime generate
the  physics of gauge theories in 4+n dimensions below their 
cut-off scale $M_s$.
\footnote{Identical 
construction has been independently given in ref.~\cite{G}, except for 
explicit discussion of the interactions between the KK modes.}
In that construction the Kaluza-Klein states of extra dimensions arise from
having many gauge symmetries in four dimensions.  The IR limit is determined
by the dynamics of spontaneous breakdown of those symmetries.

The  four-dimensional renormalizable theory constructed in
ref.~\cite{hill,G}  in the IR give the dynamics of  gauge
interactions in one extra  dimension. Let us then first recall a few 
facts about 
non-abelian $SU(n)$ gauge theories in $4+1$
dimensions.  They  have three free parameters: the cut-off scale $M_s$, the
dimensionful coupling constant 
\begin{equation}
g_0=1/\sqrt{M}\equiv g_5/\sqrt{M_s}
\end{equation}
and the compactification radius $R$ (we assume that $M_s\gg 1/R$). The 
physics seen in four dimensions is obtained by integrating out the fifth 
dimension. The simplest compactifications are on a circle $S_1$ or on an 
orbifold $S_1/Z_2$. In the latter case the four components of the vector 
potential $A_\mu(x_\mu,x_5)$, $\mu=0,1,2,3$,~ are even under $Z_2$ and the 
fifth component $A_5(x_\mu,x_5)$ is odd (here and in the following gauge 
group indices are suppressed).  Fourier expanding in the discrete fifth 
component of momentum we define the four-dimensional degrees of freedom 
(Kaluza-Klein modes): vectors $A_\mu^n(x_\mu)$, ~$n=0,1,...$, ~and scalars 
$A_5^n(x_\mu)$, ~$n=1,...$.  In the compactification on orbifold there is 
no zero mode of $A_5$ and, actually, $A_5(x_\mu,x_5)$ can be gauged away.  
If we compactify on a circle, the zero mode of $A_5$ remains in the physical 
spectrum (as a scalar in the adjoint representation of the $SU(n)$ group) 
and the massive vector KK modes are doubled. For the orbifold 
compactificaton, in the axial gauge $A_5\equiv 0$, the effective Lagrangian 
after integrating over $x_5$ contains the following well known terms:

\begin{enumerate}
\item zero mode kinetic term invariant under $SU(n)$ gauge transformations
  in four dimensions
\be
-{1\over4}F_{\mu\nu}^{(0)}F^{(0)\mu\nu}
\ee
where  
\be
F^{(0)}_{\mu\nu}=\partial_\mu A^{(0)}_\nu -\partial_\nu A^{(0)}_\mu +
\tilde{g}fA^{(0)}_\mu A^{(0)}_\nu
\ee
and $\tilde{g}=g_5/\sqrt{M_sR}$, with the coupling $g_5$ defined by eq.(1);

\item KK kinetic and mass terms
\be
\sum_{n=1}^{N}(\partial_{\mu}A_{\nu}^n-\partial_{\nu}A_{\mu}^n)^2 
+\sum_{n=1}^{N}(\frac{n\pi}{R})^2A_{\mu}^{n}A^{n\mu}  
\ee

\item triple couplings
\begin{eqnarray}
&&\frac{2g}{\sqrt{M_s R}} f \sum_{n=1}^{N} \left[
\partial_{[\mu}A_{\nu]}^0 A^{n~\mu}A^{n~\nu} + \partial_{[\mu}A_{\nu]}^n(
A^{0~\mu}A^{n~\nu} 
+ A^{n~\mu}A^{0~\nu}) \right]        \\ \nonumber 
&& + \frac{g}{\sqrt{2M_s R}}f\sum_{n,m,l=1}^{N}
\partial_{[\mu}A_{\nu]}^n A^{m~\mu} A^{l~\nu} \Delta_1(n,m,l)
\end{eqnarray}

\item quartic couplings
\begin{eqnarray}
&&\frac{g^2}{M_s R}
ff\sum_{n=1}^{N}\left(A_{\mu}^0A_{\nu}^0A^{n\mu}A^{n\nu} +
\makebox{all
permutations} \right)
\\ \nonumber 
&& +\frac{g^2}{2M_s R}ff\sum_{n,m,l,k=1}^
{N}A_{\mu}^nA_{\nu}^mA^{l~\mu}A^{k~\nu}
\Delta_2(n,m,l,k) 
\end{eqnarray}
where the $\Delta_i$ are defined as:
\begin{eqnarray}
& & \Delta_1 =
\delta(n+m-l)+\delta(n-m+l)+\delta(n-m-l)
\\ \nonumber
& & \Delta_2 = \delta(n+m-l-k)+\delta(n+m+l-k)+\delta(n+m-l+k)
\\ \nonumber
& & \qquad + \delta(n-m+l+k)+\delta(n-m-l-k)+\delta(n-m+l-k)
\\ \nonumber
& & \qquad +\delta(n-m-l+k).
\end{eqnarray}
and $f$'s are the gauge group structure constants.
\end{enumerate}

The structure of couplings reflects  conservation of  the fifth component
of momentum.  The truncation in the number of KK modes is understood,
$n<N$  where $N\approx M_sR$,  ~so that $M_N< M_s$.  
The effective Lagrangian has four-dimensional
$SU(n)$ gauge invariance,  with massive KK modes transforming linearly under
the adjoint representation of $SU(n)$,  but the full gauge invariance of
the five-dimensional lagrangian is lost because of the truncation $n<N$.
\footnote{The Lagrangian for infinite tower of KK modes has full 
five-dimensional gauge invariance realized as non-linear gauge symmetries 
for the KK modes~\cite{dienes}.} The theory is manifestly non-renormalizable.

The renormalizable four-dimensional theory that, in its infrared region,
generates  the interactions  described by the truncated Lagrangian  (i-iv)  is 
as follows~\cite{hill,G}. Let us consider the gauge structure
\begin{equation}
SU(n)_0\times SU(n)_1 \times.....SU(n)_N\equiv SU^{N+1}(n)
\end{equation}
where the vector potentials are  $A^a_{j\nu}$ and the dimensionless gauge
coupling constants are equal for all of the $SU(n)$ symmetries, 
$g_{4j}\equiv g_4$.  We suppose, in addition, that there is a set
of scalar fields $\Phi_j$ $(j=1,...N)$  (elementary or effective) which
transform as $(\bar{n},n)$ under $SU(n)_{j-1}$ and $SU(n)_j$ groups.
We shall call $\Phi_j$'s the link-Higgs fields. Their transformation 
properties define the links to be nearest-neighbour. The Lagrangian reads
\begin{equation}
{\cal L}=-{1\over4}\sum^N_{i=0} F^a_{i\mu\nu}F^{ai\mu\nu}  +
\sum^N_{i=1}(D_\mu\Phi)_i^\dagger D^\mu\Phi_i + V(\Phi_i)
\end{equation}
where $D_\mu=\partial_\mu+ig_4\sum^N_{i=0} A^a_{i\mu}T^a_i$ ~($T_i^a$
are generators of the $i$th $SU(n)_i$ gauge symmetry), and the potential 
has full chiral symmetry $SU^N(n)\times SU^N(n)$ and is symmetric
under interchange $\Phi_j\rightarrow\Phi_i$.

If the dynamics of the scalar sector (described here by the potential 
$V(\Phi_j)$; the scalars 
$\Phi_j$ can be replaced by technicolour-like condensates~\cite{G}) is such 
that the diagonal components of the scalars $\Phi_j$  acquire vacuum 
expectation value  $v$, the chiral symmetry of the link-Higgs sector is 
spontaneously broken $SU^N(n)\times SU^N(n)\rightarrow SU^N(n)$ leaving $N$
Nambu-Goldstone bosons, each transforming under the adjoint
representation of $SU(n)$.  The gauge symmetry $SU^{N+1}(n)$ of the full
Lagrangian is also spontaneously broken to the diagonal $SU(n)$ and N vector
bosons acquire masses eating up the  Nambu-Goldstone bosons.  If we want
to have four-dimensional model corresponding to a compactification on a circle,
we introduce one additional link-Higgs field $\Phi_0$ transforming as
$(3_0,\bar{3}_N)$. The chiral symmetry of the Higgs sector is now
$SU^{N+1}(n)\times SU^{N+1}(n)$. Its breaking to $SU^{N+1}(n)$ leaves $N+1$
Nambu-Goldstone bosons and $N$ of them, as before, are eaten up by longitudinal
components of $N$ vector bosons.  Thus, one massless scalar remains in the 
spectrum~\cite{wang}, in exact correspondence to the zero mode of $A_5$ in the 
5d theory.  The massless scalar transforms under the adjoint representation
of the diagonal, unbroken, $SU(n)$ symmetry.

The physics of our renormalizable 4d theory below the scale $v$ of spontaneous
gauge symmetry breaking can be easily studied in detail in the non-linear
$\sigma$-model approximation.  For the orbifold compactification, upon
substituting
\begin{equation}
\Phi_j\rightarrow v\exp(i\phi^a_jT^a/2v)
\end{equation}
the $\Phi_j$ kinetic terms lead to a mass matrix for the gauge fields

\be
\sum_{i=1}^{N} \half g_4^2v^2(A^a_{(i-1)\mu} -A^a_{i\mu})^2
\ee
This mass matrix has the structure of a nearest neighbour 
coupled oscillator
Hamiltonian. We can diagonalize the mass
matrix to find the eigenvalues :
\be
\label{Mn} 
M_n = \sqrt{2}g_4 v \sin \left[ \frac{\gamma_n}{2} \right]
\qquad \gamma_n=
\frac{n\pi}{N+1}\ , \qquad n=0,1,\dots, N. 
\ee
Thus we see that for small $n$ this system has a KK tower
of masses given by:
\be
M_n \approx   \frac{g_4 v\pi n}{\sqrt{2}(N+1)}\qquad \qquad n\ll N
\ee
and $n=0$ corresponds to the zero-mode. 

To match on to the spectrum of the KK modes, we require 
\be
\frac{g_4 v}{\sqrt{2}(N+1)} = \frac{1}{R}. 
\ee
Hence, the system with $SU(n)^{N+1}$ and $N$ $\Phi_i$ provides 
a  description of the  KK modes of the 5d theory by generating the same 
mass spectrum. 

Next, it is  crucial to examine the interactions from the
model. 
The gauge fields $A_{\mu}^{j}$ can be expressed as linear combinations of the
mass eigenstates $\tilde{A}_{\mu}^{n}$ as: 
\be 
A_{\mu}^j = \sum_{n=0}^{N} a_{jn} \tilde{A}_{\mu}^n~. 
\label{AA}
\ee
The $a_{nj}$ form a normalized
eigenvector ($\vec{a}_{n}$)
associated with the $n$th $n \neq 0$ eigenvalue 
and has the following components:
\be
a_{nj} =\sqrt{{2\over N+1}}\cos\left(\frac{2j+1}{2}\gamma_n\right)\ , \qquad
j = 0, 1,\dots, N, 
\label{anj}
\ee
The eigenvector for the zero-mode, $n=0$ ,
is always $\vec{a}_{0} =
\frac{1}{\sqrt{N+1}} (1,1,\dots,1)$. 
We can now rewrite the Lagrangian eqn.(9) in the mass eigenstates of
the vector bosons ($\tilde{A}_{\mu}^{n}$) and derive the interactions between
them. They are identical to those of the 5d theory, eq.(3-6), with the 
identification
\be
\tilde{g}=\frac{g_5}{\sqrt{M_sR}}=\frac{g_4}{\sqrt{(N+1)}} 
\ee

In the 4d  theory,  there are three relevant parameters, namely, the 
gauge coupling
constant $g_4$, the total number of $SU(3)$ groups $N+1$ and the VEV of the
Higgs field $v$ determined by the potential $V(\Phi_i)$. The mappings between 
them and the parameters of the 5d
theory are $N+1 =
M_s R$, $g_4=g_5=\sqrt{M_s/M}$ and $v=\sqrt{M_sM}$.

Another interesting fact is that
the 4d model can also be interpreted as the 5d theory on  a
``transverse lattice'' \cite{transverse}.
We construct a transverse lattice in the $x^5$ dimension
where the lattice size is $R$ and short-distance lattice cut-off
is $a=1/M_s$, so $N+1 =R/a$. This
is a foliation of $N+1$ parallel branes, each spaced
by a  lattice cut-off $a$. 
On the $i$th brane
we have an $SU(n)$ gauge theory denoted by $SU(n)_i$.  The
$SU(n)_i$ automatically have a common 
coupling constant $g_4$.

The theory thus
has $N$ links in the $x^5$ direction that are
continuous functions of $x_\mu$. These
correspond to the continuum limit Wilson lines:
\be
\Phi_i(x^\mu) = \exp\left[ig_4\int_{ia}^{(i+1)a}\; dx^5 A_5(x^\mu, x^5)\right]
\rightarrow \exp\left[ig_4a A_5\left(x^\mu, (i+\half)a\right)\right]
\ee
The $N$ $\Phi_i$ therefore transform as an $(\bar{n},n)$ representation
of $SU(n)_i\times SU(n)_{i+1}$ as in the 4d model.
$\Phi_i$ is a unitary matrix and may be parameterized 
as in eq.(10).

The construction of ref.~\cite{hill,G} provides new insights and opens
up several interesting possibilities.  The most straigthforward one is 
to discuss the 5d physics in the framework of a 4d renormalizable field 
theory. For instance, the power-law like running of the gauge couplings 
in 5 dimensions \cite{tom,dienes} can be discussed in the renormalizable 
setting. In the 4d theory, the running is logarithmic but with the $\beta$ 
function changing every time we pass the threshold $M_n$ \cite{hill}. In 
the limit of large $N$ one recovers the power-law like running.

Another obvious implication of these considerations is that the transverse
lattice is {\it an exact} approach to 5d gauge theories with a cut-off.

Furthermore, one can use five-dimensional models as inspiration for 
four-dimensional theories and vice versa. Following the first direction,
new ideas have been proposed on electroweak symmetry breaking\cite{arkani,
jing} and on supersymmetry breaking~\cite{grojean,cheng} mechanisms in
four dimensions.  Going in the reversed direction, one can get new insight
into fermion mass generation in five dimensions~\cite{wang}.
A link to string theory is discussed in ref.~\cite{sf}.

Finally and perhaps most interestingly,
the theory constructed in ref.~\cite{hill,G}
offers an alternative, to the geometrical one, interpretation of gauge fields
propagating in the bulk as arising from nearest-neighbour 
coupled, spontaneously broken, extended gauge symmetries in four dimensions.
The big question is if similar interpretation can be extended to gravitational 
interactions. If so, one would get a new view on the notion of dimension.

\vskip1.0cm

\noindent {\bf Acknowledgments}
I thank P.~Chankowski, J.P.~Derendinger, E.~Dudas, A.~Falkowski, Ch.~Grojean, 
Z.~Lalak, H.P.~Nilles  and G.~Ross for interesting conversations.
This work was supported partially by the Polish State Committee for Scientific
Research grant 5 P03B 119 20 for 2001-2002 and by the EC Contract 
HPRN-CT-2000-00152 for years 2000-2004.


\begin{thebibliography}{99}
\bibitem{hill} C.T.~Hill, S.~Pokorski, J.~Wang, hep-th/0104035 (to be 
               published in {\em Phys. Rev.} {\bf D}).

\bibitem{G} N.~Arkani-Hamed, A.G.~Cohen and H.~Georgi, 
            {\em Phys. Rev. Lett.} {\bf 86}, 4757 (2001), hep-th/0104005.

\bibitem{wang} Hsin-Chia Cheng, C.T.~Hill, S.~Pokorski, J.~Wang, preprint
               FERMILAB-PUB-01-053-T, hep-th/0104179 
               (to be published in {\em Phys. Rev.} {\bf D})

\bibitem{transverse} W.A.~Bardeen, R.B.~Pearson and E.~Rabinovici,
                     {\em Phys. Rev.}  {\bf D21}, 1037 (1980).

\bibitem{tom} T.R.~Taylor and G.~Veneziano, {\em Phys. Lett.}
              {\bf B212}, 147 (1988).
 
\bibitem{dienes} K.R.~Dienes, E.~Dudas and T.~Gherghetta,
                 {\em Nucl.\ Phys.} {\bf B537}, 47 (1999); 

\bibitem{arkani} N.~Arkani-Hamed, A.G.~Cohen and H.~Georgi, preprint 
                 HUTP-01-A024, hep-ph/0105239.

\bibitem{jing} H.-C.~Cheng, Ch.T.~Hill and J.~Wang, preprint 
               FERMILAB-PUB-01-079-T, hep-ph/0105323.

\bibitem{grojean} C.~Csaki, J.~Erlich, Ch.~Grojean and 
                 G.D.~Kribs, preprint UCB-PTH-01-18, hep-ph/0106044.

\bibitem{cheng}  H.-C.~Cheng, D.E.~Kaplan, M.~Schmalz  
                 and W.~Skiba, preprint FERMILAB-PUB-01-108-T, hep-ph/0106098.

\bibitem{sf} K.~Sfetsos, preprint NEIP-01-005, hep-th/0106126.

\end{thebibliography}
\end{document}